\documentclass[%
reprint,
 amsmath,amssymb,
 aps,
prl,
]{revtex4-2}

\usepackage{graphicx}
\usepackage{dcolumn}
\usepackage{bm}
\usepackage{color}
\usepackage{soul}
\usepackage{hyperref}

\begin{document}

\preprint{APS/123-QED}

\title{Two relaxation mechanisms for rejuvenation of stable polystyrene glass}

\author{Saba Karimi}
\author{Junjie Yin}

\author{James A. Forrest}%
 
\affiliation{Department of Physics and Astronomy, University of Waterloo, Waterloo, Ontario, Canada}

\date{\today}

\begin{abstract}
We report on the rejuvenation of thin films of polystyrene (PS) as they are heated from stable glassy states - prepared either through vapour deposition or  physical aging.  For films with thickness $h \gtrsim 200$ nm, the rejuvenation of vapour deposited stable PS glass films follow behaviour well-documented for other stable glasses.  For films with thickness $h \lesssim 160$nm  the behaviour of the vapour deposited films becomes more complicated.  This behaviour is characterized for different film thicknesses.  The results are compared with vapour deposited films that are aged after rejuvenation or spincast and aged. Collectively, the results of these studies strongly suggest two distinct mechanisms that can result in rejuvenation, and hence two distinct relaxation processes that can couple to the material density.  The two physical mechanisms can provide insight into poorly understood or apparently contradictory aspects of $T_g$ measurements in thin PS films.  \end{abstract}

\maketitle

\section{Introduction}

Finite size effects are a leading class of experiments to find measures of a physical length scale associated with cooperative motion in glass formers.  One way to study dynamics is to exploit the connection between dynamics and measures of the glass transition temperature  $T_g$. The study of glass transitions in nanoconfined geometry can be traced back to the work of  Bares for polymer systems ~\cite{bares1975glass} and later to Jackson and McKenna for organic liquids confined to pores ~\cite{jackson1991glass}. In those studies it was found that the  $T_g$ value measured by calorimetry was reduced below the bulk value when materials were confined to nanometric dimensions.  A breakthrough in finite size experiments came with the work of Keddie, Jones and Cory ~\cite{keddie1994size} where they used ellipsometry to measure the dilatometric  $T_g$ values for thin films of polystyrene. The measured $T_g$ values were found to be reduced below the bulk value for films with thickness $h$ less than $\sim$ 50 nm.  This work attracted significant attention, and measurements of the glass transition in thin polymer films has been an  active research area for three decades since that first report~\cite{forrest2001glass,roth2005glass}.  While many different polymers have been measured, the majority of the work has involved polystyrene, and the effect in PS is among the largest for any material. These effects have been correlated to and proposed to be caused by  enhanced mobility in the near surface region~\cite{sharp2004properties,forrest2014does,sharp2003free}. 

The free surface of many glassy materials has been shown to exhibit non-glassy behaviour~\cite{tian2022surface}.  In fact the surface of glassy materials can exhibit significant mobility more characteristic of a liquid than a glass~\cite{chai2014direct,yin2023surface}. This has been thought of, and quantitatively described as either a liquid-like layer resulting from a strong mobility gradient, or as surface diffusion.  With the description of a liquid-like layer on top of a bulk glassy solid, it has been demonstrated that the measured $T_g$ value in thin films are determined by the condition that half of the film has the liquid value of the thermal expansivity.  This already makes thin film measurements of $T_g$ inherently different than a bulk material where each part of the material (suitably averaged over time) behaves the same. Thus, in a bulk material held at $T_g$, all parts of the sample will have a characteristic alpha relaxation time of about 100s.  In a thin film held at it's $T_g$ parts of the film near the surface will have a characteristic alpha relaxation time of much less than 100s, and parts of the film further away from the free surface will be characterized by relaxation times that are many orders of magnitude larger. Thus the thin film $T_g$ measurements are not providing the exact same measure as they do in bulk materials.  The length scale of the penetration of the surface dynamics into the rest of the material is correlated to the length scale of dynamic cooperativity.  Such an approach has been seen to offer a quantitative description of measured $T_g$ values in thin PS films~\cite{forrest2014does,salez2015cooperative}.  The link between reduced $T_g$ values as measured by dilatometry and enhanced surface mobility was brought into question by measurements of Paeng {\em et al}~\cite{paeng2011direct}. Those experiments showed no evidence for a layer of enhanced segmental mobility (anywhere) in poly ($\alpha$ methyl styrene) with $M_w$ of 488 kg/mol. At the same time Geng {\em et al} measured significant reductions in $T_g$ of films with thickness less than $h \sim 40$nm of  poly ($\alpha$ methyl styrene) films with $M_w$ value of 20, and 420 kg/mol but no reductions for $M_w$ of 1.3 kg/mol thickness ~\cite{geng2015molecular}.  

Significant theoretical advances have recently been made in understanding the underlying cause of the structural glass transition in infinite dimensions~\cite{parisi2020theory}. Application of these ideas to  measurable dynamics on real systems remains an active area. In particular the realization (or existence) of an ideal glass, and the measurement of a length scale over which structural relaxation is cooperative remain open questions. These two problems relate to two fields of experimental efforts.  The existence of ideal glasses is encompassed in studies of recently discovered ultrastable glasses produced by physical vapour deposition, and the the idea of a length scale for cooperative motion is the driving force for measurements of the glass transition in nanoconfined geometries. 

The observation of ultrastable glasses prepared by Physical vapour deposition was first reported by Swallen {\em et al.}~\cite{swallen2007organic}. Since that time, many groups have prepared and characterized stable glass materials. This has been accompanied by computer simulations studies as well~\cite{berthier2017origin}.  Vapour deposition leading to stable glasses has been measured in molecular liquids, metals, and polymers.  Authoritative reviews of this field can be found in  ~\cite{rodriguez2022ultrastable,ediger2017perspective}.
Stable glass formation is thought to result from the same enhanced surface mobility that has been linked to reductions in $T_g$ value~\cite{berthier2017origin}.  The link between the dynamics of the supercooled liquid and the velocity of front propagation provides another way to probe dynamics in thin films of glass formers~\cite{Karimi2024measurement}. This provides  the added advantage of having depth resolution for films where the rejuvenation proceeds by front propagation.

In this work, we investigate the rejuvenation of stable glasses of PS prepared by physical vapour deposition or by physical aging of (spincast and vapour deposited) films over a range of thicknesses.  A number of unexpected and surprising observations are made.  Explanation of these observations requires postulation of two distinct physical relaxation mechanisms. The presence of two different relaxation mechanisms can also impact how we understand thin films glass transition measurements in polystyrene. While we are not able to provide evidence that unequivocally confirms our ideas, their potential impact on thin film glass transitions is significant.

\section{Experimental details}
Vapour deposited polystyrene stable glass films are prepared as described previously ~\cite{raegen2020ultrastable}.  The films used in this study have $N\sim 9$ and $T_g$ values between 302.7 K and 314.7 K, and have $T_f/T_g$ values from 0.93 to 0.97. Ellipsometric measurements are made either  with a Film-sense FS-1 Multi-Wavelength ellipsometer or a J.A. Woollam M-2000 spectroscopic ellipsometer. Ellipsometric models are used to convert these measured quantities to film thickness and refractive index of the constituent layers. For these PS materials, the contrast in refractive index between the stable glass and the supercooled liquid is small (fully stable and  fully rejuvenated have  $\Delta n \sim  0.01$). For most studies we employ a single layer model to provide an average thickness and refractive index. Spincast films with the same range in thickness as those produced by vapour deposition were prepared from toluene solutions of the same pre-distilled polymer samples.  A key difference between vapour deposited films and spincast films- even using the same distilled material- is that the process of vapour deposition favours the lowest polymer index in the source and materials produced this way have small average $N$ values for the first depositions, and are much more monodisperse. As more depositions are performed with the same source, the $N$ values available become larger, and the $T_g$ values continually increase. Originally spincast films are aged at a temperature of 293.2 K and have a measured $T_g$ of 318.2 K.  

\begin{figure*}
\centering
\includegraphics[width=0.9\textwidth]{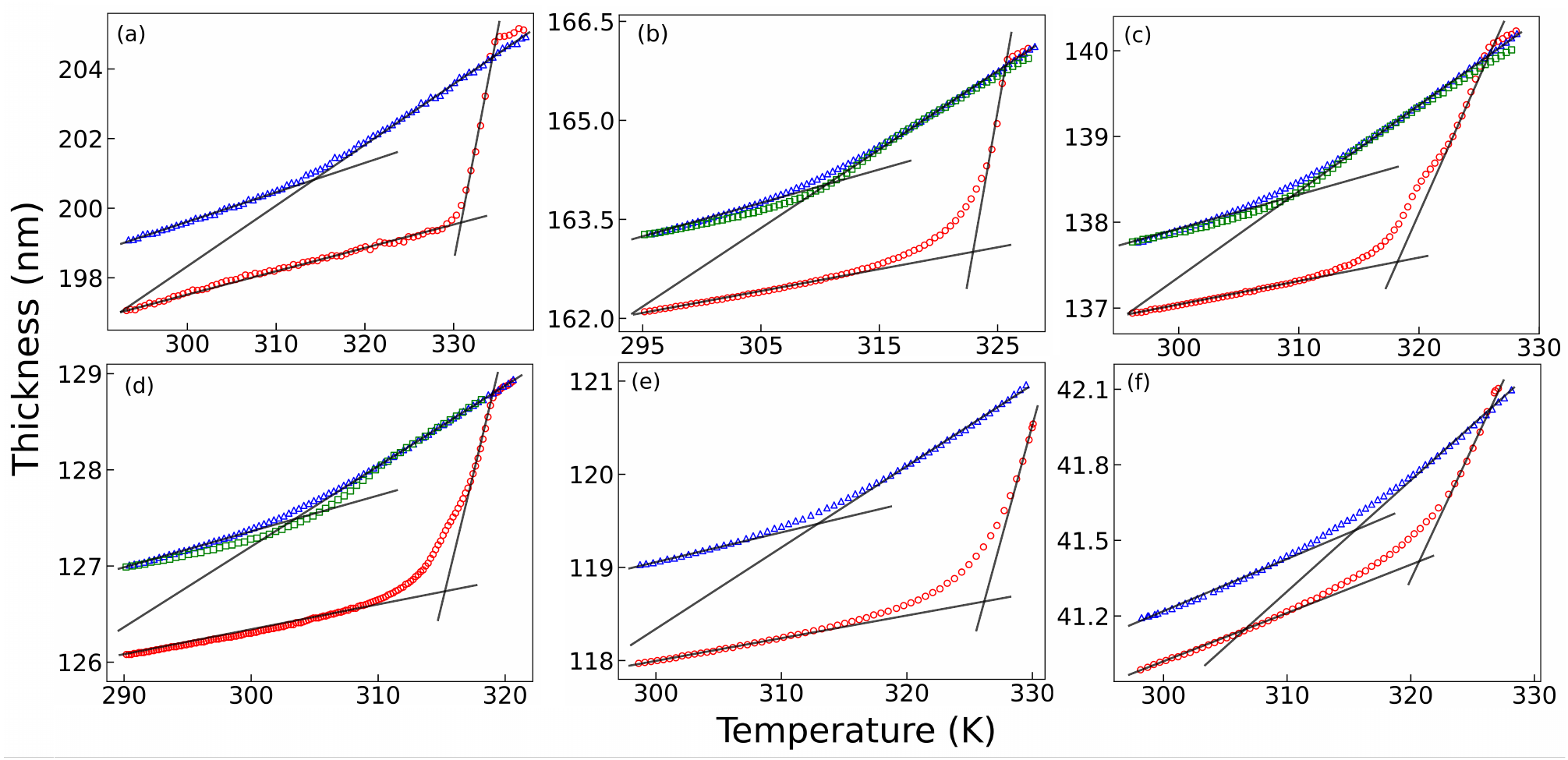}
\caption{ Film thickness obtained from single film modelling of ellipsometric data for the first heating (red circles) and first cooling (blue triangles) for thick, intermediate and thin films of stable PS glass produced by vapour deposition.   In some cases the data from a second heating immediately after the first cooling is also shown (green squares).}
\label{fig1}
\end{figure*}

\section{Results and Discussion}

Fig.1 shows a series of thickness versus temperature graphs (heat/cool) for a series  of  vapour deposited PS  glasses.   Fig 1(a) shows the  first heating of a thick film ($\gtrsim 200$ nm ) from the stable glass state. This is the same general behaviour that is reported for molecular stable glasses as well on first heating.   In the stable glass state, the material exhibits a thermal expansion value that is similar to, but usually less than, the freshly cooled glass. This continues until a temperature referred to as the onset temperature, $T_{ons}$, at which the material rapidly undergoes a transformation to a supercooled liquid. This dynamical process leads to a very large apparent value of $\frac{dh}{dT}$ . For thin films, this transformation occurs through propagation of a front that originates at one or both of the interfaces ~\cite{flenner2019front,walters2015thermal,dalal2015influence}.  By definition, the value of $T_{ons}$ is always greater than $T_g$ for a kinetically stable glass. After this transformation is complete, the $\frac{dh}{dT}$ is  the thermal expansion of the equilibrium supercooled liquid. Upon cooling,  the supercooled liquid will simply retrace the liquid heating behaviour (as expected from an equilibrium liquid)  and will continue to exhibit this thermal expansivity until it reaches the $T_g$ value where the slope will change over a small interval of temperature to a smaller value characteristic of the glass.  This first heating and cooling curve is used to determine the fictive temperature $T_f$, the onset temperature $T_{ons}$, and the glass transition temperature $T_g$ as described in numerous studies ~\cite{ediger2017perspective,Dalal2013HighthroughputEC}. The thermodynamic stability can be characterized by the ratio $\frac{T_f}{T_g}$ and the kinetic stability by the value $\frac{T_{ons}}{T_g}$.   

Fig. 1 demonstrates the evolution of the film thickness dependence of rejuvenation for glasses with similar stability (as quantified by the value of $\Delta \rho/ \rho$). For films with a thickness $h \lesssim 200$nm changes in the  thickness versus temperature behaviour during the first heating scan begin to manifest.    One of the most obvious differences for the thinner films is that there are changes in the thickness versus temperature plot at temperatures far below the eventual measured $T_{ons}$.   The first deviation of the expansivity from that of the stable glass (in panels b-f)  seems to occur at  $T \sim T_g$.  Upon heating to $T>T_g$, the expansivity  increases rapidly but continuously.  Film thickness dependent changes in the response to temperature change can be further resolved.  In particular, for a small thickness range near $h \sim 140$ nm there is an apparent two step rejuvenation.  Similar anomalous behaviour (including non monotonic changes) of the thickness versus temperature on first heating within the context of a single layer model have been reported in other systems ~\cite{dalal2012molecular}. Such changes can be attributed to the failure of the single layer model being used to describe what is essentially a 2 layer system upon rejuvenation.  In such cases, the large differences in refractive index between the stable glass layer and the rejuvenated liquid layer can result in a sample that is not well modelled by a one layer approach. Such large differences in refractive index can be caused by molecular ordering that can occur during vapour deposition and has been measured in a number of molecular stable glasses ~\cite{dalal2015tunable}. This is unlikely to be the explanation for these PS films for two reasons.  First the refractive index difference between the stable and supercooled layer is much less than in many molecular systems. Second, this apparent two step behaviour is observed in only a very small thickness interval. For the current case of oligomeric/polymeric materials the ordering can involve the side groups (in this case the phenyl rings) and can be decoupled from ordering of the chain backbone. Previous ellipsometric measurements  in stable glass of PS for films in the range 100 to 800 nm did not provide any evidence of birefringence, although it is hard to correlate this to an upper bound for the amount of molecular orientation ~\cite{raegen2020anisotropy}.

\begin{figure*}
\centering
\includegraphics[width=0.3\textwidth]{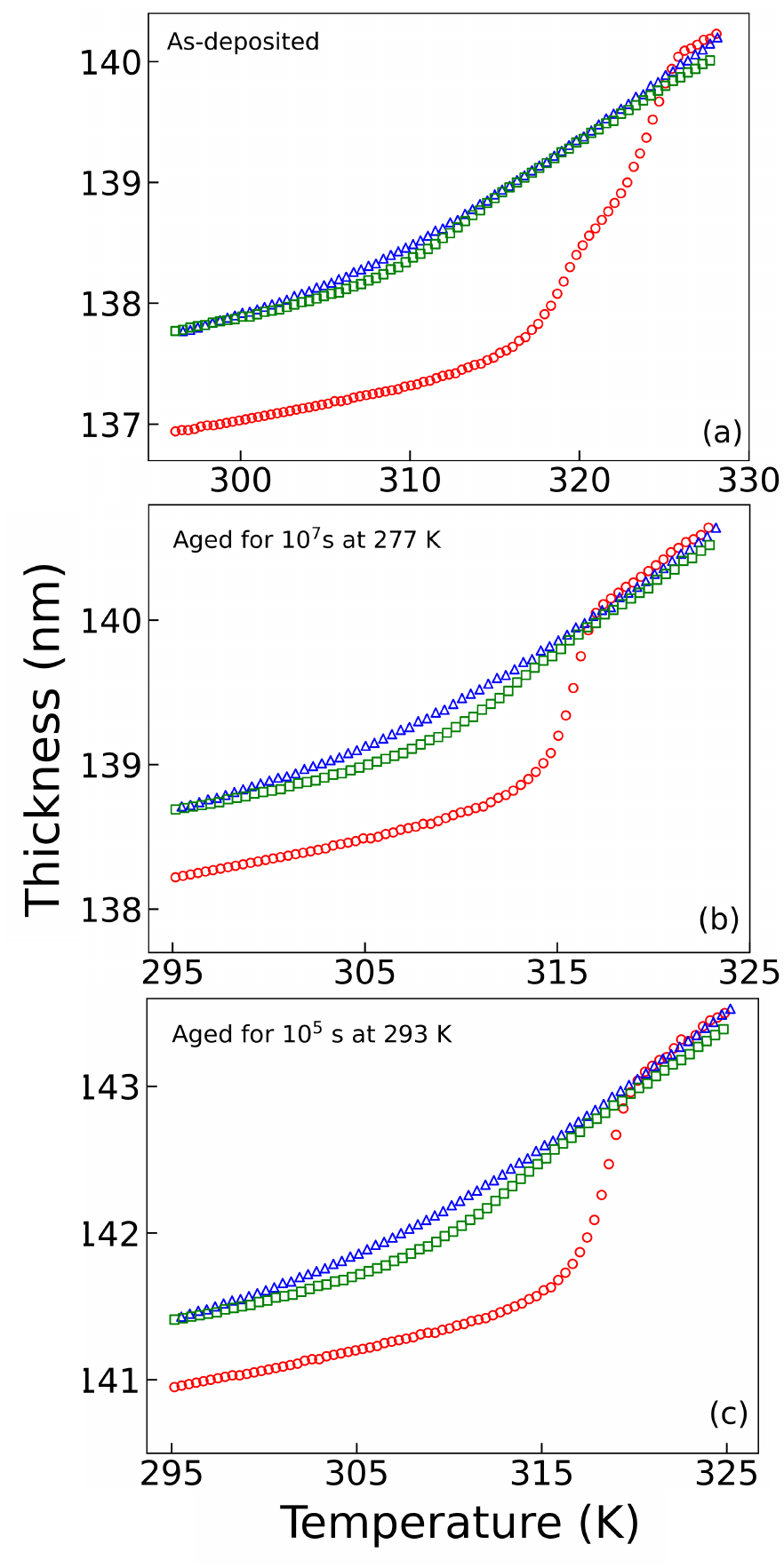}
\caption{ Film thickness versus temperature obtained from single film modelling of ellipsometric data for the first heating (red circles) and cooling (blue triangles) and second heating (green squares) showing anomalous behaviour over very small range in film thickness. Fig. 2(b) is aged for $10^7$ s at 277K, and Fig. 2(c) is aged  for $10^5$ s at 293 K.}
\label{fig2}
\end{figure*}

 In order to investigate whether the step rejuvenation is more likely a result of two distinct relaxation processes or simply due to inadequacies of single layer modelling, we can consider films that have been aged after the first heat/cool cycle. This produces films which are stable, are the identical polymer material, and have the identical thickness of the vapour deposited sample.  If the apparent two step behaviour is due to inadequacies of the one layer modelling without some changes in the molecular ordering, then aged films with the identical thicknesses should  also show the steps.  On the other hand, if the sample produced by vapour deposition has an alternative relaxation route, then aged and vapour deposited films with the same apparent stability will evolve differently on rejuvenation.  Fig. 2 shows a single sample measured right after vapour deposition as well as the same sample after a heat cool cycle and then being aged at 293 K for 10 days or 277 K for 3 months. In the as-deposited sample there are two apparent processes with different onset temperatures. One process has an onset temperature of $T_g+14$ K, and the other has an onset temperature of $T_g+10$ K. Each of the processes appear to involve a similar stability in terms of $\frac{\Delta h}{h}$. In both cases the aged samples show no evidence for the two step rejuvenation suggesting it is not due to modelling.  The onset temperature in the aged glass is $T_g+5$ K, but it is hard to directly compare the values of  $T_{ons} $ as the vapour deposited sample are generally  analogous to samples with much longer aging times. The lack of a second process and a second $T_{ons} $ in the aged rather than vapour deposited samples suggests that vapour deposition allows for a second stability mechanism that is less  stable than the one obtained solely by aging. This idea is supported by the observation that vapour deposited films with this thickness that have been allowed to age for many months after deposition do not exhibit the first rejuvenation step (data not shown).  In such samples, it is likely the rejuvenation of that less stable process is able to occur during the aging time. As a second test, we  produce aged films of the same material that are produced by spincasting and allowed to age for 10 days. By producing many films over a small range of thickness we can investigate whether such aged films show the two step behaviour seen for vapour deposited films.  By making a series of 4  different film thicknesses in the range 120 to 140 nm, we can ensure that some of the films will be in the range coinciding with the two step process in as-deposited vapour deposited films.  In no case do the aged spincast films show the same two step relaxation, or any anomalous structure in the rejuvenation plot. We note that the aged samples have a stability (determined by $\frac{\Delta \rho}{\rho}$) which is roughly half that of the vapour deposited samples and the same as either of the two apparent rejuvenation  processes measured for the as-deposited sample. We also note that the thickness region where two step relaxation is observed appears to be a function of the polymerization index of the material, but this dependence not been thoroughly characterized. In the current study we are comparing materials within a narrow range of  polymerization index. 

\begin{figure*}
\centering
\includegraphics[width=0.4\textwidth]{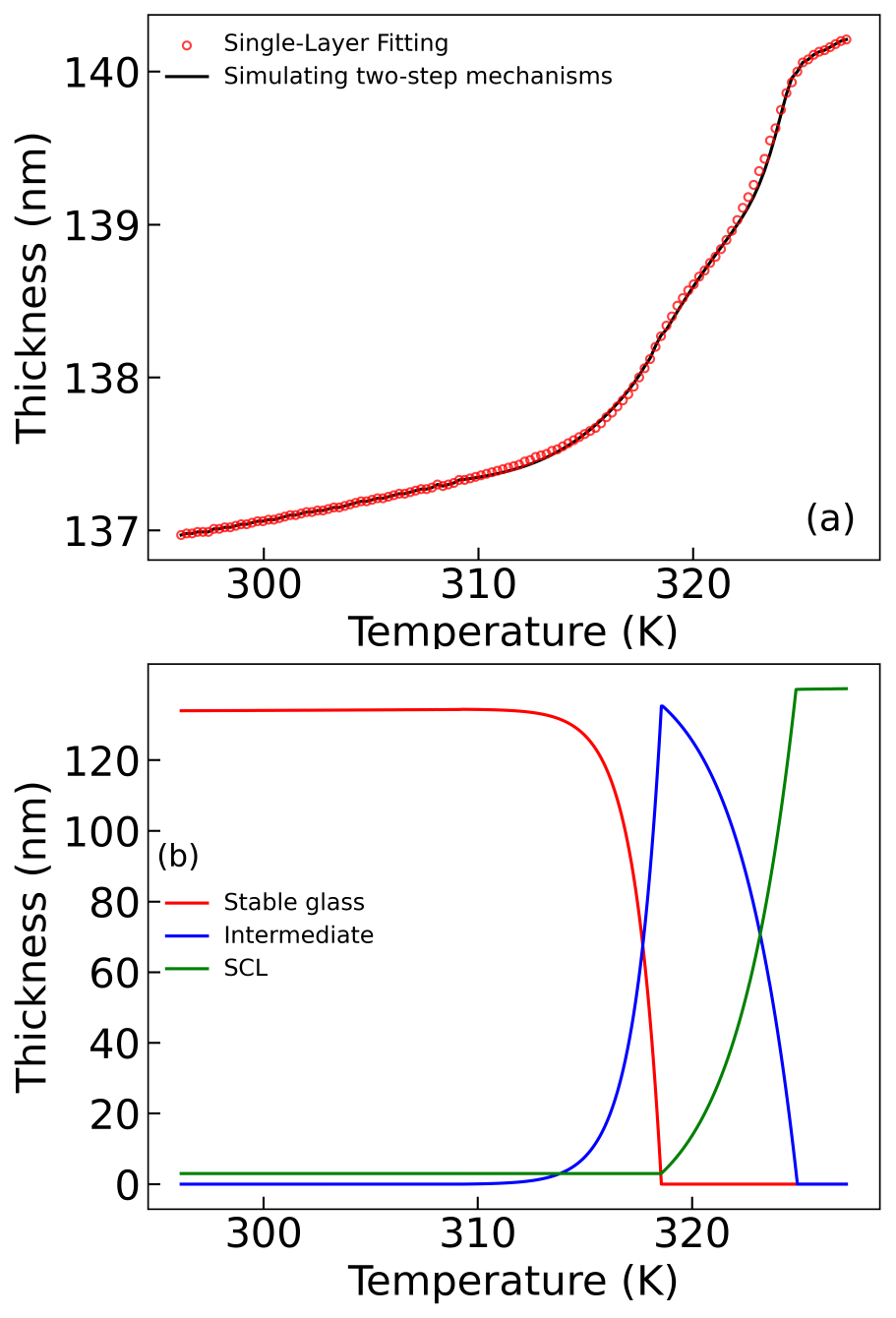}
\caption{(a) Film thickness obtained from single film modelling of ellipsometric data for first heat and cool cycle (symbols) as well as that obtained by fitting ellipsometric data obtained from three state model to single film modelling. (b) Thickness of stable glass, intermediate state, and supercooled liquid as a function of temperature used to generate the solid curve in (a). Details of the modelling are described in the text. }
\label{fig3}
\end{figure*}

We can quantify the behaviour  by considering the idea of two different mechanisms of stability and modelling the rejuvenation accordingly.  In this process we have three materials, the stable glass, the rejuvenated supercooled liquid, and an intermediate material where one mechanism causing stability has been rejuvenated, but the other has not. This intermediate state could be one where initially ordered side groups have become disordered while maintaining the higher density of the backbone segments. This physical picture requires three thickness values (stable glass, intermediate material, supercooled liquid) and three refractive index values. In order to allow a tractable comparison, we fix the temperature dependent refractive index of the stable glass and supercooled liquid phases by the high temperature and low temperature extreme behaviour of the sample.  Further, we assume that the refractive index of the intermediate material can be approximated as $n_{int}(T)=\frac{1}{2}(n_{stable}(T)+n_{liquid}(T))$ (as suggested by the fact that each process appears to involve half of the total value of $\frac{\Delta \rho}{\rho}$). We model the temperature dependence of the system by using the fact that the front velocity increases exponentially with increasing temperature ~\cite{Karimi2024measurement}. This gives an exponential relation between front position and temperature.  By assuming that the first mechanism rejuvenates completely with increasing temperature before the second front begins propagating, we can numerically produce a thickness versus temperature for each layer. We then calculate the resulting ellipsometric parameters of the simulated sample using the matrix approach of Hayfield and White ~\cite{hayfield1964ellipsometry} to compare to experimental data. Varying the exponent in the temperature dependent rejuvenation and the temperature where each mechanism starts, we can generate an equivalent one film description that can be compared with the date shown in Fig.2(a). The results of this process are shown in Fig. 3. It is clear from this Figure that a description where a first front passes through the material and relaxes one process followed by a second front relaxing the main process is able to provide a quantitative description of the data. 

\begin{figure*}
\centering
\includegraphics[width=0.7\textwidth]{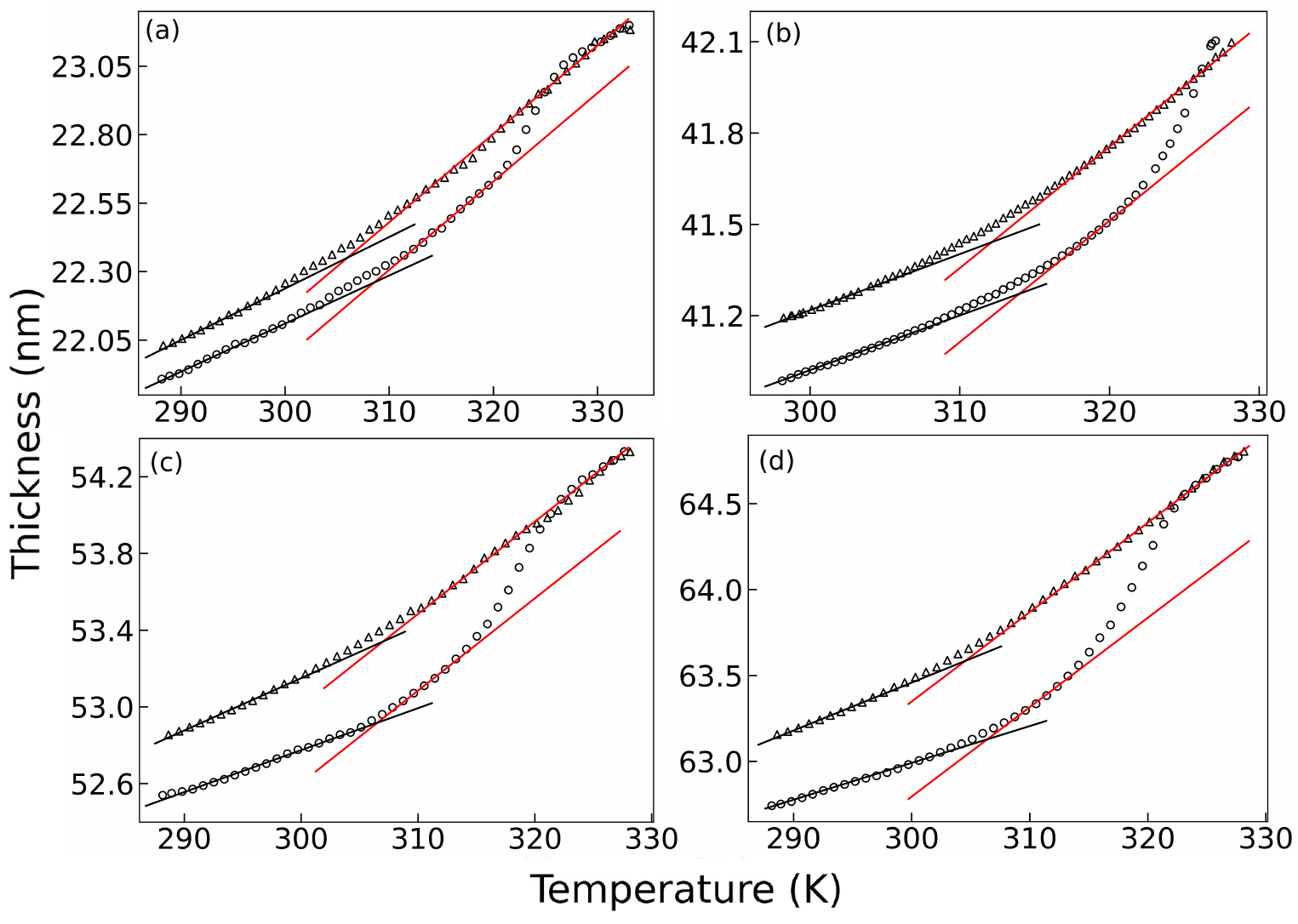}
\caption{Film thickness obtained from single film modelling of ellipsometric data for the first heating and cooling for very thin films ($h \lesssim 50$ nm) produced by vapour deposition showing apparent liquid-like expansivity over an extended temperature range. Red lines have slopes defined by the liquid expansivity.}
\label{fig2}
\end{figure*}

For films with thickness $h\lesssim 60$ nm, the thickness versus temperature behaviour is remarkable and qualitatively different from both  the thick films and of the behaviour described above. The thin film behaviour is shown in Fig. 4.  In this case there is still a first deviation of glassy expansivity at the $T_g$ value.  Instead of increasing to large values characteristic of rejuvenation in temperature ramps, for an extended range of temperatures, $T>T_g$ but $<T_{ons}$ the film displays a thermal expansivity which is the same as that of the liquid even though the material still has a thickness less than that of the liquid cooled glass. In Fig. 4 both straight red lines are fits to the supercooled liquid expansivity with one being simply offset to compare to the expansivity in the stable glass.   The material at $T < T_{onset}$ is obviously still kinetically stable as further heating above $T_{ons}>T_g$ leads to rejuvenation of the material into the ordinary liquid, and yet this stable glass has a thermal expansivity the same as the supercooled liquid.   The onset of this intermediate behaviour occurs at the bulk $T_g$ value.  Similar deviations from the stable glass expansivity at $T \sim T_g$ are also evident in molecular glasses~\cite{jin2021glasses}.  One might be tempted to attribute this  behaviour as being due to a part of the film becoming normal and exhibiting liquid expansivity. However if that were the case, the slope of the $h$ versus $T$ data would be intermediate between the liquid and the glass. Instead, the slope is always the same as that measured eventually for the equilibrium liquid. In fact  in the interval  $T>T_g$ but $<T_{ons}$ the data is quantitatively described as the entire film expanding with the value of the liquid expansivity. We can summarize these measurements by saying that between the $T_g$ value of the freshly cooled glass, and the $T_{ons}$, the thin film has a thermal expansion coefficient the same as the equilibrium liquid (thereby suggesting it is a liquid) while at the same time has a density greater than the equilibrium liquid at that temperature. It is also the case that any time the $\frac{dh}{dT}$  changes from the glassy value to the very large apparent values characteristic of rejuvenation kinetics there will necessarily be some temperature interval where the dh/dT value is the same as the equilibrium liquid.  Notably,  it is {\em only in this thin film region} that we see an extended range of temperatures where the value of $\frac{dh}{dT}$ is quantitatively described by the liquid value of the thermal expansivity. 

\begin{figure*}
\centering
\includegraphics[width=0.4\textwidth]{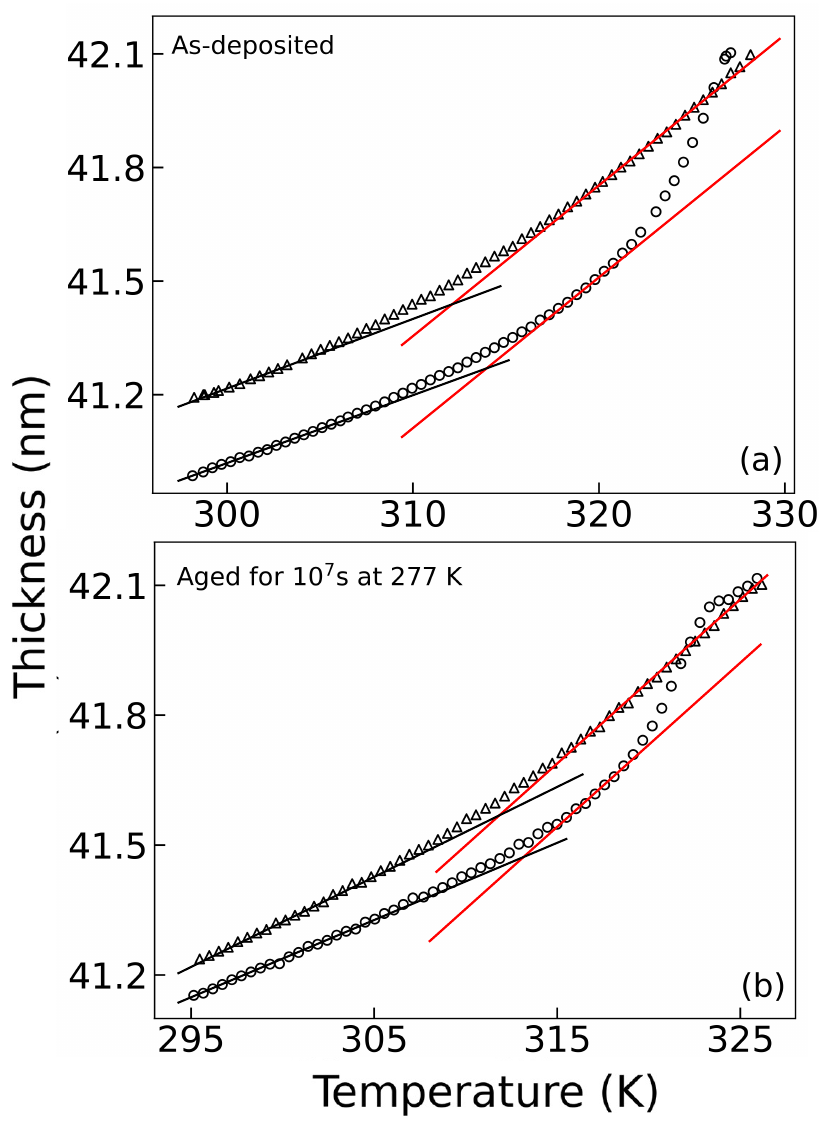}
\caption{ Film thickness obtained from single film modelling of ellipsometric data for heating and cooling for very thin films ($h \lesssim 50$ nm) which were produced by vapour deposition, then rejuvenated and allowed to age at 277K for $4 \times 10^7$s. }
\label{fig2}
\end{figure*}

We also consider samples that are vapour deposited, rejuvenated, and then allowed to age. Figure 5 shows a comparison between a film on first heating after vapour deposition and the same film after aging 15 months at 277 K.  The behaviour is quantitatively very similar. Both samples show an extended temperature range where  the sample is obviously stable and  yet exhibits a $\frac{dh}{dT}$ that is the same as the equilibrium liquid. The only quantitative difference is that the  as-deposited version  has a $T_{ons}$ of $\sim 322$ K, and the rejuvenated then aged version of the same film has a  $T_{ons} $ of $\sim 317$ K. This indicates that while the behaviour is qualitatively the same, the vapour deposited sample is (as expected) more kinetically stable than the sample aged for 15 months.  

Similar to the case for intermediate thicknesses, the thickness regime where this behaviour is observed is a function of the polymerization index, $N$.  This  thin film rejuvenation behaviour also suggests the possibility of two distinct rejuvenation processes which may then indicate the possibility of two distinct relaxation mechanisms that can couple to the film density. We note that other studies have also suggested more than one relaxation mechanism observable in thin films of polystyrene.  Svanberg ~\cite{svanberg2007glass} noted the existence of a long length scale relaxation in dielectric studies of free standing films of PS. Yuan {\em et al} has similarly noted an additional  long length scale process in mechanical relaxation measurements in thin films~\cite{yuan2022microscale}.  Recent studies on physical aging in polystyrene films have also been used to  support the idea of two different relaxation mechanisms~\cite{SHARMA2021124103}. 

The types of motions normally discussed in terms of glass forming materials can be roughly broken up into two main categories.  The main alpha relaxation is comprised of relaxation processes that allow different configurations to be sampled.  This is usually the motion that is most strongly coupled to the material density. In contrast more local processes do not allow different configurations to be sampled but may still couple to the material density through different packing due to molecular orientation. In small molecules, these local motions could be motions within the nearest neighbour cage, and for polymers they are associated typically with the motion of side groups. For the case of polystyrene, it is most likely that the local mechanism would involve some small number of  phenyl groups, and the longer length scale mechanism would relate to cooperative relaxation of at least a few segments along the backbone. If each type of motion can couple to the material density then each mechanism can also contribute to the thermal expansion.   Normally (but not necessarily) the main alpha relaxation has the largest impact on both the thermal expansion and the heat capacity. Upon cooling below $T_g$,  the segmental relaxation becomes frozen, but the local  processes remain active. The local process may be frozen out at  still lower temperatures.  

A decoupling between the main alpha relaxation and the majority of the  thermal expansion has been suggested for measurements of poly ($\alpha$ methyl styrene). In that material, measures of the segmental mobility in free standing glassy films showed no evidence for a liquid like layer ~\cite{paeng2011direct}.  This was in contrast to many other polymer materials which showed strong evidence for a layer of highly mobile material (presumably near the free surfaces). This lack of mobility is in contradiction of measurements of reduced $T_g$ values in thin films of poly ($\alpha$ methyl styrene) ~\cite{geng2015molecular} within the well accepted framework that $T_g$ reductions are a result of enhanced dynamics in the near free surface region. Paeng {\em et al} ~\cite{paeng2011direct} suggested that dilatometric $T_g$ was not necessarily an accurate reporter of segmental mobility in free-standing films. " While the correlation between $T_g$ and segmental mobility is very well established in bulk polymers, it is possible that the inference that the segmental relaxation time is $\sim 100$ s at $T_g$ is not correct in thin free-standing films." For the case of polystyrene the importance of local phenyl motions in determining the calorimetric or dilatometric $T_g$ was also discussed in dynamic infrared linear dichroism measurements of PS ~\cite{noda2009glass}. Those studies suggested that the normally quoted $T_g$ was mainly associated with local phenyl ring motions, and larger collective backbone motions was not evident until the temperature reached $\sim T_g+25$ K. Notably, this contrasts the observation that  thin films showed evidence of whole chain flow at the dilatometric $T_g$ ~\cite{chai2014direct}. The relevance of phenyl motion can be compared to  measurements of phenyl-reorientation in rubbed glassy films. Tsang {\em et al} ~\cite{tsang2001temporal} noted that relaxation of rubbed films is orders of magnitude faster than segmental motions. In such samples, relaxation was observed at temperatures as low as 60 K below the $T_g$ value. The relaxation of phenyl orientation was described by a sun of two exponentials with one having a lifetime of $\sim 10^2$ s, and the other having a lifetime of $\sim 10^4$ s within the temperature range $T_g$ - 60 K to $T_g$ -10 K. 

With these studies in mind we suggest the rejuvenation results above arise from the existence of two different physical processes. One physical process is the collective motion of segments usually associated with the glass transition, and the other is a more local process likely dominated by phenyl ring reorientation. Each of these physical processes can couple to the material density.  This is similar to the recent suggestion of Luo {\em et al} ~\cite{luo2024length} that two physical processes contribute to aging in PS glasses. Within these two processes we note that the motion of segments is obviously a sufficient but not necessary condition for phenyl reorientation. In order to explain the rejuvenation results for thin films, it is necessary to postulate that a significant component of the liquid thermal expansion is due to phenyl reorientation even in the absence of segmental mobility.  With this idea, phenyl reorientation even with the backbone motion still frozen,  can lead to liquid-like expansivity. We further suggest that the free surface provides kinetic facilitation for the phenyl reorientation in some region near the free surface. Note that the ability of phenyl rings on polystyrene to orient near the free surface may be related to the significant enhancements of aging on thin films of toluene ~\cite{sepulveda2011accelerated}. The anomalous rejuvenation behaviour over a very small range of thickness near 140 nm can be due to the vapour deposition process being able to produce a sample with an orientation of phenyl rings which results in higher overall density.  As the temperature is raised, the phenyl ring ordering relaxes first, followed by the main chain relaxation. Upon aging this sample again, the phenyl ordering can not occur and re-aging of the rejuvenated film only results in the stability of the segmental motion.

This  explanation suggested to explain the observed rejuvenation  has implications for measurements of $T_g$ in thin films being reduced below the bulk value.  The current measurements show that it is possible that {\em there is a film thickness dependent mechanism that allows  liquid-like thermal expansion even when the film is in an ultrastable glass state}.  This clearly demonstrates the possibility of measuring an apparent glass transition in the absence of segmental mobility.  For example in Fig. 4, we could easily have defined  a $T_g$ by the the change in expansivity that happens for $T \sim T_g$, but that material for $T>T_g$ is still a stable glass, and not a liquid.  We can be sure of this because as temperature is further raised we can see the transformation into the liquid.  After transformation the material has the same liquid like expansivity as for $T_g<T<T_{ons}$. While somewhat uncommon, the idea of having causes for steps in the thermal expansion in PS films is not new.  Roth {\em et al} measured two $T_g$ values in free standing films of PS~\cite{pye2011two}.  Dalnoki-Veress {\em et al}~\cite{dalnoki2000glass}, and and Mattsson {\em et al}~\cite{mattsson2000quantifying} noted that in high $M_w$ free standing films that exhibit large changes in apparent $T_g$, the films can be held for seemingly arbitrarily long times at $T>T_g$ without showing the hole formation that is evidence of flow.  In fact this stability is what makes the larger reductions in apparent $T_g$ possible to measure However, in these same samples once the temperature is raised to $T \sim T_g(bulk)$ hole formation and growth due to flow rapidly occurs. Recent theoretical work provides strong evidence that the $M_w$ dependence of $T_g$ in free standing PS films can not be related to the main alpha relaxation process ~\cite{bonneau2023bridge}. All of these results in free standing films of PS are consistent with the step in expansivity at lower temperatures not being coupled to segmental motion, and being instead related to this other mechanism involving phenyl motion which has been noted both in this rejuvenation of stable glass, and aging in PS glass.  

If this interpretation is correct then both the currently reported two mechanisms for rejuvenation, similar reports of two aging mechanisms ~\cite{luo2024length} and the decades old report of $M_w$ dependent reductions in thin free standing films of PS may share a common underlying origin. In the case of rejuvenation of thin stable glasses, and measurements of the $T_g$ in thin films, the liquid-like thermal expansion is due to a local motion rather than longer length scale segmental mobility.  In both cases, that motion is facilitated by either segmental motion or release of constraints at the free surface.  The effect of the free surface should have some propagation length, and thus we expect and observe similarities between the thickness range where these two effects are observed.  Interestingly, the use of a kinetic facilitation model has also been used to lead to a logarithmic dependence of the $T_g$ on the $M_w$ ~\cite{baker2022cooperative}.  This is the same puzzling (and so far unexplained) dependence used to quantify the  measured apparent $T_g$ values in thin free standing films with $M_w \gtrsim 500$k.

We have introduced the idea that a liquid like value of the thermal expansion in PS does not necessarily mean the material is not in the glassy state. The fact that two $T_g$ values have been measured~\cite{pye2011two} and inferred ~\cite{mattsson2000quantifying} suggests that step changes in the expansivity in thin films in some cases still do correlate to reductions in the $T_g$ due to enhanced segmental mobility.   It appears that it is non-trivial to distinguish changes in expansivity due to local motions or real segmental motion in very thin films of PS. In such cases, measurements of $T_g$ should be accompanied by some measure of the dynamics (in the current case the transformation to the normal liquid).   While it is not possible yet to prove the ideas beyond any doubt, the ideas provide a mechanism to explain the current puzzling observations, as well as provide insight into measurements which have been at the centre of controversy for 3 decades now. 

\section{Acknowledgements}
The authors are grateful to the Natural Science and Engineering Research Council (NSERC) of Canada for supporting this research.  J.Y. acknowledges support from Waterloo Institute for Nanotechnology through a WIN Nanofellowship.

\bibliographystyle{ieeetr}
\bibliography{references}

\end{document}